\documentclass[aps,superscriptaddress,amsmath,amssymb,floatfix,twocolumn,showpacs,amsfonts,longbibliography]{revtex4-1}
\usepackage{times}
\usepackage[varg]{txfonts}
\usepackage{textcomp}
\usepackage{graphicx}
\usepackage{subfigure}
\usepackage{tabu}
\usepackage{color}
\usepackage[colorlinks=true,citecolor=blue,urlcolor=blue,linkcolor=blue,hyperindex]{hyperref}
\usepackage{braket}
\usepackage{overpic}
\usepackage{amssymb}
\allowdisplaybreaks

\begin{document}
	\preprint{Physical Review Letters}
	
	\title{Dynamical spin excitations of topological Haldane gapped phase in the $S=1$ Heisenberg antiferromagnetic chain with single-ion anisotropy}

	\author{Jun-Han Huang}
	\affiliation{State Key Laboratory of Optoelectronic Materials and Technologies, School of Physics, Sun Yat-Sen University, Guangzhou 510275, China}
	
	\author{Guang-Ming Zhang}
	\email{gmzhang@tsinghua.edu.cn }
	\affiliation{State Key Laboratory of Low-Dimensional Quantum Physics and Department of Physics, Tsinghua University, Beijing 100084, China}
	\affiliation{Frontier Science Center for Quantum Information, Beijing 100084, China}
	
	\author{Dao-Xin Yao}
	\email{yaodaox@mail.sysu.edu.cn}
	\affiliation{State Key Laboratory of Optoelectronic Materials and Technologies, School of Physics, Sun Yat-Sen University, Guangzhou 510275, China}
	
	\date{\today}
	
	\begin{abstract}
		We study the dynamical spin excitations of the one-dimensional $S=1$ Heisenberg antiferromagnetic chain with single-ion anisotropy by using quantum Monte Carlo simulations and stochastic analytic continuation of imaginary-time correlation function. Using the transverse dynamic spin structure factor, we observe the quantum phase transition with a critical point between the topological Haldane gapped phase and the trivial phase. At the quantum critical point, we find a broad continuum characterized by the Tomonaga-Luttinger liquid similar to a $S=1/2$ Heisenberg antiferromagnetic chain. We further identify that the elementary excitations are fractionalized spinons.
	\end{abstract}

	\maketitle

	\section{INTRODUCTION}
\begin{figure}[t]
		\centering
		\includegraphics[width=8cm]{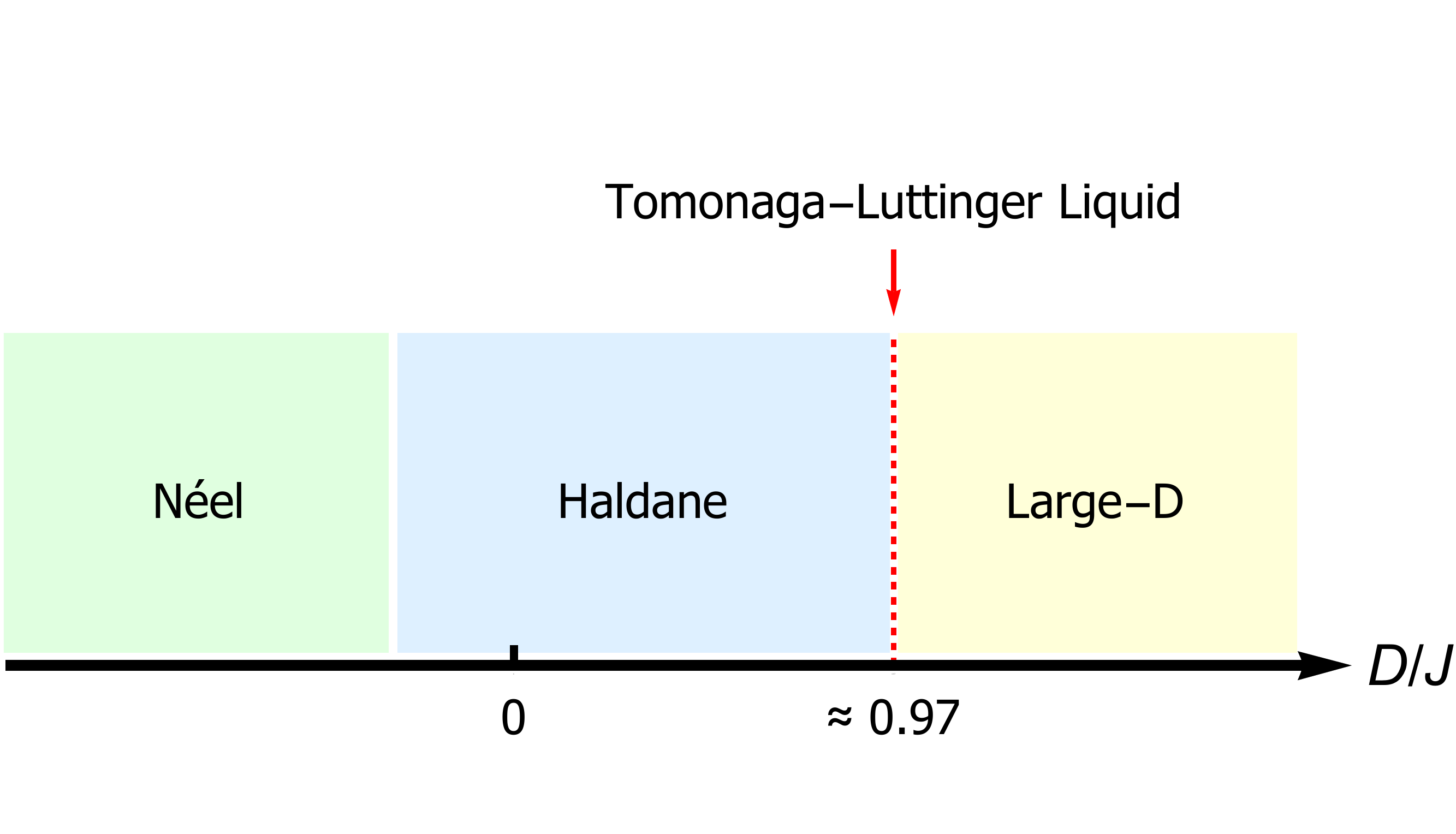}
		\caption{Quantum phase diagram of the $S=1$ Heisenberg chain versus single-ion anisotropy $D/J$. The quantum critical point between the Haldane phase and the large-D phase is a Tomonaga-Luttinger liquid phase (red dashed line).}
		\label{phase}
	\end{figure}

	An extensive body of research has been done on the one-dimensional $S=1$ Heisenberg antiferromagnetic chain, which can be traced back to the original work by Haldane \cite{PhysRevLett.50.1153,PhysLettA93.464}. Here we investigate the spin chain with single-ion anisotropy \cite{PhysRevB.97.060403} and present the quantum phase diagram \cite{PhysRevB.79.054412,PhysRevB.67.104401,PhysRevB.77.224435} with a Ising-type Néel phase, a Haldane phase and a large-$D$ phase in Fig. \ref{phase}. The phase transition from the Néel order to the Haldane phase has been studied in detail before \cite{PhysRevB.99.045117}, which is described by conformal field theory with a central charge $c=1/2$ \cite{PhysRevA.77.012311}. Thus, we are more interested in the other transition from the Haldane phase to the large-$D$ phase, i.e. a continuous transition. This quantum phase transition is a Gaussian-type transition \cite{PhysRevB.84.220402} described by conformal field theory with a central charge $c=1$ \cite{PhysRevB.90.075151,EurPhysJB35.465}. The Haldane gapped phase is now understood as a symmetry-protected topological phase \cite{PhysRevB.93.165135}, while it will undergo a phase transition to a topologically trivial gapped phase, the large-$D$ phase, with increasing the single-ion anisotropy $D$. When $D$ is strong enough, the ground state in the large-$D$ phase is known as the product state with $S^z=0$ at each site. A symmetry-protected topological phase cannot be continuously changed into a trivial gapped phase without closing the energy gap \cite{PhysRevLett.111.080401}, so the transition, between the Haldane phase and the large-$D$ phase, is a topological phase transition that possesses a gapless critical point and does not fit into the Landau-Ginzburg-Wilson paradigm \cite{PhysRevB.70.144407}.

	The previous research \cite{PhysRevB.91.054405} has suggested that a direct transition from the Haldane phase to the trivial phase can occur without accessing a Tomonaga-Luttinger liquid (TLL) critical state in the absence of external magnetic field, which is inaccurate, or rather easy to be neglected. Recently, the TLL phase of the $S=1/2$ chain has been realized experimentally by the rare-earth perovskite $\mathrm{YbAlO}_3$ \cite{NatCommun10.698} and exhibited a broad continuum, a signature of fractionalized spinon excitations, as predicted by various theoretical and numerical methods \cite{PhysRevB.98.174421,PhysRevLett.122.175701,PhysRevLett.121.117202,PhysRevLett.98.227202}. This provides strong evidence to identify that the critical point separating the Haldane phase from the trivial phase is a TLL critical state by the spin excitation spectra theoretically and experimentally. The $S=1$ chain with single-ion anisotropy can be realized by ultracold atomic condensates on optical lattices and various compounds with $\mathrm{Ni}^{2+}$ ions, such as $\mathrm{Ni}(\mathrm{C}_2\mathrm{H}_8\mathrm{N}_2)_2\mathrm{NO}_2(\mathrm{ClO}_4)$ (NENP), $\mathrm{NiCl}_24\mathrm{SC}(\mathrm{NH}_2)_2$ (DTN) and so on \cite{PhysRevB.69.020405,PhysRevB.50.9174,EurophysLett3.945,PhysRevLett.96.077204}. In addition, experimental methods, including inelastic neutron scattering and nuclear magnetic resonance, provide the dynamic probes of these materials \cite{NatPhys11.62}.

	\begin{figure*}[t]
		\centering
		\includegraphics[width=18cm]{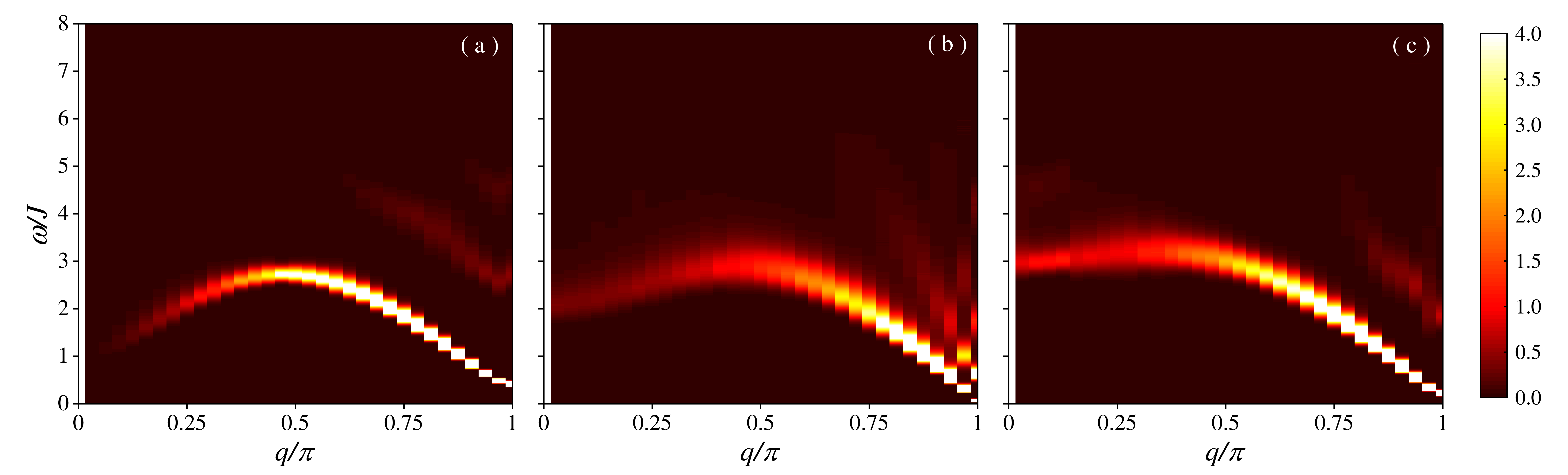}
		\caption{The transverse dynamic spin structure factor $S^{xx}(q,\omega)$ obtained from QMC-SAC calculations for the $S=1$ Heisenberg chain with $L=64$ and $\beta=128$. The values of single-ion anisotropy are (a) $D/J=0$ in the Haldane phase, (b) $D/J=0.97$ in the TLL critical state and (c) $D/J=1.5$ in the large-D phase, respectively.}
		\label{sqw}
	\end{figure*}

	In this paper, we study the dynamics of the one-dimensional $S=1$ Heisenberg antiferromagnetic chain with single-ion anisotropy \cite{PhysRevB.78.174429}. By means of quantum Monte Carlo (QMC) \cite{Sandvikbook,PhysRevB.66.024418,PhysRevB.61.364,PhysRevE.66.046701,PhysRevB.66.134407} simulations and stochastic analytic continuation (SAC), we present and discuss our results for the transverse dynamic spin structure factor. The QMC-SAC numerical methods provided excitation spectra very well in previous studies \cite{PhysRevLett.120.167202,PhysRevB.97.104424,PhysRevB.99.085112}. Here, we study the dynamical spin excitations of the topological Haldane phase and its phase transition to the topologically trivial large-$D$ phase. At the quantum critical point, we show the comparison with the TLL phase of the $S=1/2$ Heisenberg antiferromagnetic chain.

	\section{MODEL AND NUMERICAL METHODS}
	
	\subsection{Model}
	We investigate the anisotropic $S=1$ Heisenberg antiferromagnetic chain defined by the Hamiltonian
	\begin{equation}
	\begin{aligned}
	H=\sum_i[J(S_i^{x}S_{i+1}^{x}+S_i^{y}S_{i+1}^{y}+S_i^{z}S_{i+1}^{z})+D(S_i^{z})^2],
	\end{aligned}
	\label{Hamiltonian}
	\end{equation}
	where $S_i^{x,y,z}$ denotes the $S=1$ spin operator on each site $i$, $J>0$ is the antiferromagnetic exchange, and the parameter $D$ is the single-ion anisotropy. For simplicity, we set $J=1$ in the whole paper.
	
	As shown in Fig. \ref{phase}, the phase diagram of this model consists of the Néel phase, the Haldane phase, the TLL critical state and the large-$D$ phase. For the isotropic case, i.e. $D/J=0$, the ground state belongs to the symmetry-protected topological Haldane phase with a Haldane gap of $\Delta\approx0.41J$ \cite{PhysRevLett.69.2863,PhysRevB.48.3844} and the lowest-lying excitations is magnon. Whereas at finite $D$, the magnon excitations will split into a singlet branch ($S^z=0$) and a doublet branch ($S^z=\pm1$), which show up in the longitudinal and transverse dynamic spin structure factors respectively \cite{PhysRevB.48.311}. In the paper, we are more interested in the phase transition from the Haldane phase to the large-$D$ phase, in which the lowest-lying excitations lie in the $S^z=\pm1$ branch.
	
	\subsection{Methods}
	We numerically solve the model in Eq. (\ref{Hamiltonian}) by using QMC simulations based on the stochastic series expansion \cite{PhysRevB.59.R14157,PhysRevE.64.066701}. Using stochastic analytic continuation of imaginary-time correlation function obtained from QMC simulations, we extract the transverse dynamic spin structure factor $S^{xx}(q,\omega)$, which is written in the basis of eigenstates $\ket{n}$ and eigenvalues $E_n$ of the Hamiltonian as
	\begin{equation}
	\begin{aligned}
	S^{xx}(q,\omega)=\pi\sum_n|\langle n|S_q^x|0\rangle|^2\delta[\omega-(E_n-E_0)].
	\end{aligned}
	\label{dynamic}
	\end{equation}	
	Here, the momentum-space operator $S_q^x$ is the Fourier transform of the real-space spin operator
	\begin{equation}
	\begin{aligned}
	S_q^x=\frac{1}{\sqrt{L}}\sum_le^{-iql}S_l^x,
	\end{aligned}
	\label{fourier}
	\end{equation}
	where $q=2n\pi/L$, $n=1,2,…,L$ for periodic boundary condition. Thus, we can study the dynamics of magnetic materials naturally by the transverse dynamic spin structure factor $S^{xx}(q,\omega)$, which is convenient to compare with experiment.
	
	The stochastic series expansion QMC algorithm is used to compute the imaginary-time correlation function
	\begin{equation}
	\begin{aligned}
	G_q^{xx}(\tau)=\langle S_{-q}^x(\tau)S_q^x(0)\rangle.
	\end{aligned}
	\label{imaginary}
	\end{equation}
	And its relationship to the transverse dynamic spin structure factor is
	\begin{equation}
	\begin{aligned}
	G_q^{xx}=\frac{1}{\pi}\int_{-\infty}^{\infty}d\omega S^{xx}(q,\omega)e^{-\tau\omega}.
	\end{aligned}
	\label{relation}
	\end{equation}
	The imaginary-time correlation function $G_q^{xx}(\tau)$ can be calculated by the spectral function $S^{xx}(q,\omega)$. However, the inverse process is hard to solve because of statistical error and non-uniqueness. In SAC, we propose a candidate spectral function from the Monte Carlo process and fit them to the imaginary-time data according to a likelihood function
	\begin{equation}
	\begin{aligned}
	P(S)\propto\exp\left(-\frac{\chi^2}{2\Theta}\right),
	\end{aligned}
	\label{likehood}
	\end{equation}
	where $\chi^2$ is the goodness of fit and $\Theta$ is the sampling temperature. Finally, we can obtain the optimal spectra through such Metropolis sampling algorithm. A more detailed account of SAC can be found in Refs. \cite{PhysRevE.94.063308,PhysRevLett.86.528,PhysRevX.7.041072,PhysRevB.57.10287,PhysRevB.79.094409}.

	\begin{figure}[t]
		\centering
		\includegraphics[width=8cm]{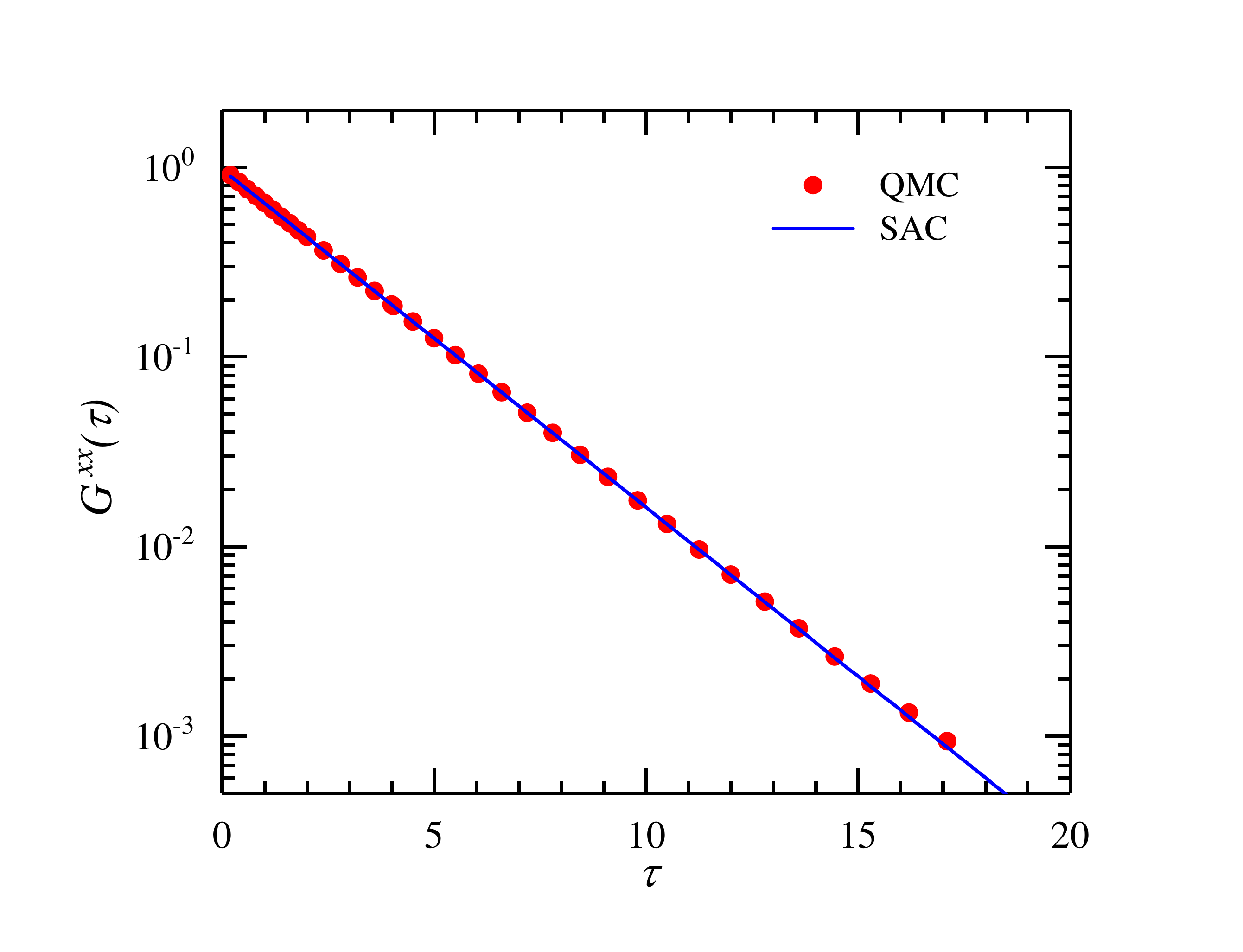}
		\caption{(Color online) The normalized imaginary-time correlation function $G^{xx}(\tau)$ at the wave vector $q=\pi$ for the $S=1$ Heisenberg chain with $D/J=0$, $L=64$ and $\beta=128$, computed in stochastic series expansion QMC calculations. The error bars are much smaller than the symbols. The straight line (blue solid line) corresponds to the contribution from the single-magnon peak obtained from SAC, with the amplitude $a_0=0.97375(25)$ and the energy gap $\Delta\approx0.4104J$.}
		\label{gtau}
	\end{figure}
	
	\section{NUMERICAL RESULTS}
	Here, we consider a $S=1$ Heisenberg antiferromagnetic chain of $L=64$ with periodic boundary condition and the inverse temperature $\beta=1/T=2L$ unless specifically mentioned. For positive $D$, the lowest-lying excitations are extracted in the transverse dynamic spin structure factor. In the paper, we study the transverse dynamic spin structure factor $S^{xx}(q,\omega)$ of the topological Haldane phase and its phase transition to the topologically trivial large-$D$ phase.
	
	\subsection{Transverse dynamic spin structure}
\begin{figure}[t]
		\centering
		\includegraphics[width=8cm]{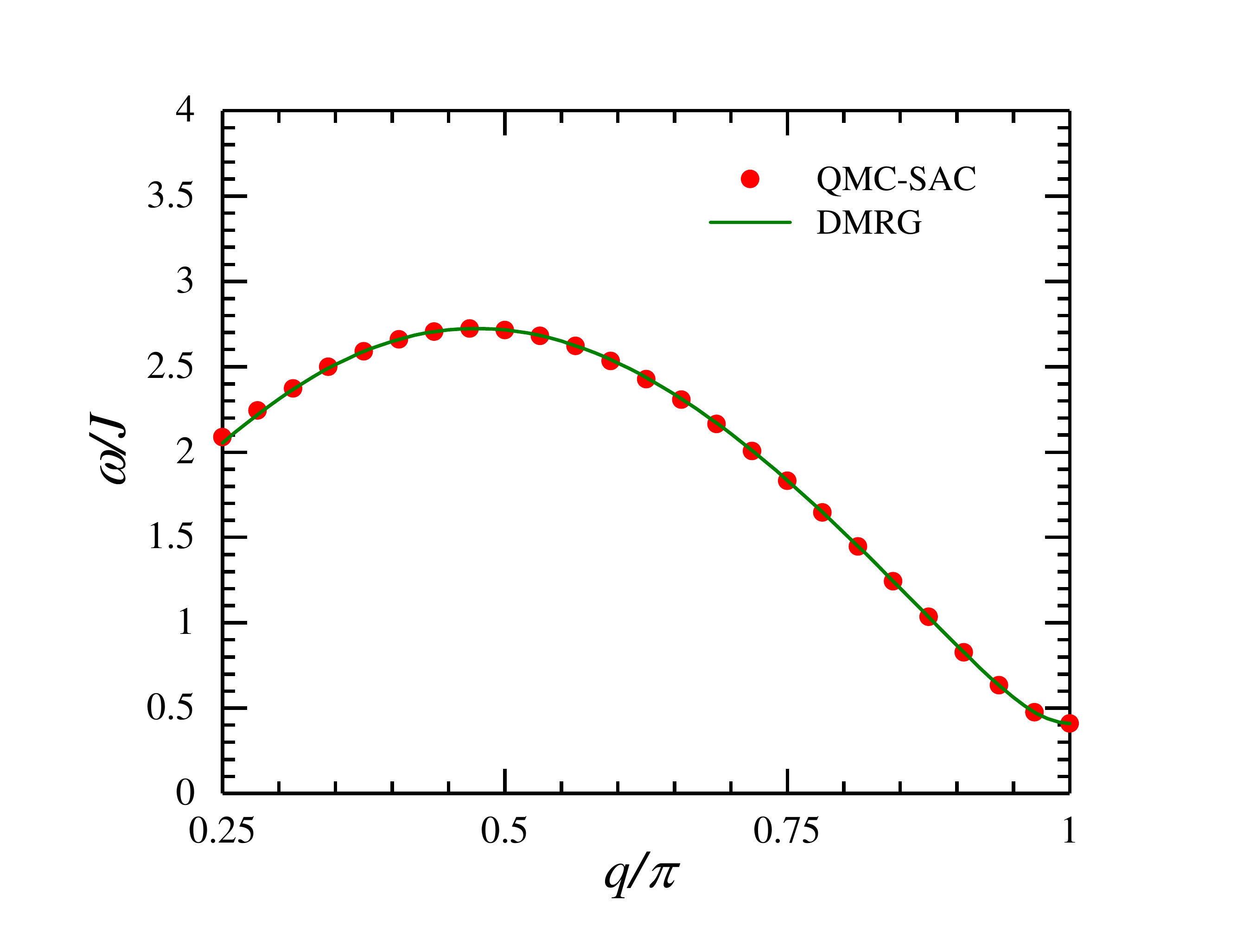}
		\caption{(Color online) The single-magnon dispersion. The red circles show QMC-SAC data for single-magnon peak ($\delta$ function) at the lower frequency bound. The curve (green solid line) is the DMRG results according to Ref. \cite{PhysRevB.77.134437}.}
		\label{sqw0}
	\end{figure}

	In the Haldane phase \cite{PhysRevB.96.060403,ComMathPhys147.431,PhysRevB.94.214427}, we assume an isotropic case, i.e. $D/J=0$. In Fig. \ref{sqw}(a), we show the results of the transverse dynamic spin structure factor $S^{xx}(q,\omega)$ obtained from QMC-SAC calculations. The most prominent contribution to excitation spectra is the single-magnon peak, with a lowest energy gap of $\Delta\approx0.41J$ at the wave vector $q=\pi$. The energy gap $\Delta$ in our methods is given simply by the imaginary-time correlation function
	\begin{equation}
	\begin{aligned}
	G_{q=\pi}^{xx}(\tau)\approx a_0e^{-\Delta\tau},
	\end{aligned}
	\label{gtaugap}
	\end{equation}
	where $a_0$ is amplitude of the single-magnon peak. As shown in Fig. \ref{gtau}, the transverse imaginary-time correlation function $G^{xx}(\tau)$ obtained from stochastic series expansion QMC calculations is exponential decay at the wave vector $q=\pi$. It is easy to extract the energy gap $\Delta\approx0.41J$ from the fitting to Eq. (\ref{gtaugap}). The excitation spectra is a single-magnon peak followed by extremely weak multimagnon continua at higher frequencies, so we provide special treatment to the single-magnon $\delta$ function at the lowest frequency when optimizing the candidate spectral function \cite{PhysRevX.7.041072}. In SAC, we also obtain that the amplitude of the single-magnon $\delta$ function is $a_0=0.97375(25)$ and its energy gap is $\Delta\approx0.4104J$ with higher accuracy. The contribution of the single-magnon excitation obtained from SAC nearly coincides with the real transverse imaginary-time correlation function $G^{xx}(\tau)$ as shown in Fig. \ref{gtau}. The single-magnon peak of momenta $q$ is displayed in Fig. \ref{sqw0}. The results obtained from QMC-SAC are perfectly consistent with the previous DMRG results \cite{PhysRevB.77.134437}. Thus, our numerical methods and results are accurate and reliable enough. 
	
\begin{figure}[t]
		\centering
		\includegraphics[width=8cm]{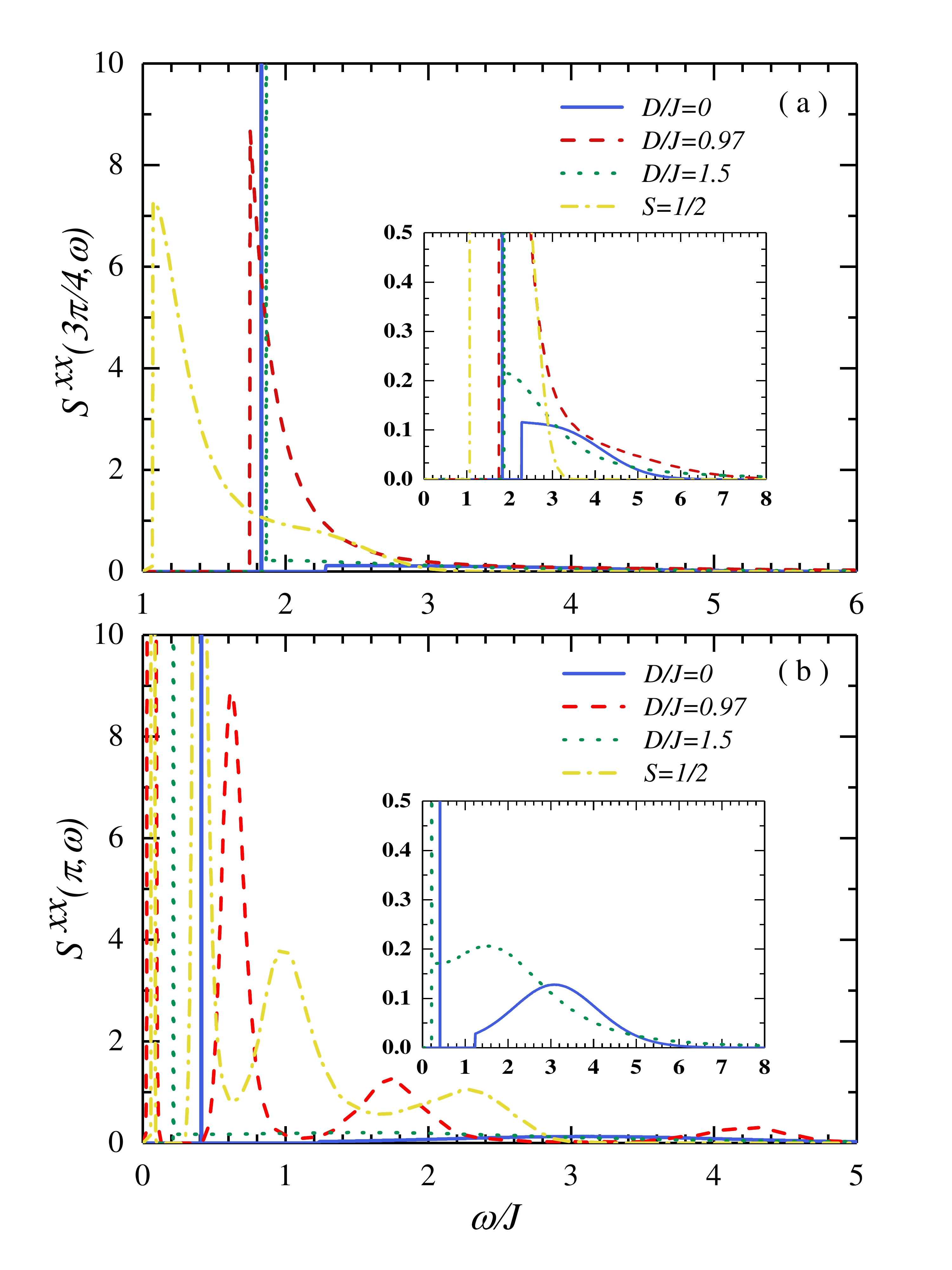}
		\caption{(Color online) The transverse dynamic spin structure factor $S^{xx}(q,\omega)$ of the $S=1$ and $S=1/2$ chains with $L=64$ and $\beta=128$ at two momenta, (a) $q=3\pi/4$ and (b) $q=\pi$. The $S=1$ Heisenberg chain is calculated with three different anisotropies, including $D/J=0$ (blue solid line), $D/J=0.97$ (red dashed line) and $D/J=1.5$ (green dotted line). The blue solid line and green dotted line both have a very sharp peak. The red dashed lines, at the quantum critical point, possess a broad continuum similar to the $S=1/2$ chain (yellow dot-dash line).}
		\label{spiw}
	\end{figure}

	In addition, in Fig. \ref{sqw}(a), the two-magnon and three-magnon continua can also be observed near the wave vectors $q=0$ and $q=\pi$, respectively, although their spectral weights are very small. The inset of Figure \ref{spiw}(b) presents the three-magnon continuum at $q=\pi$ starting at higher frequency $3\Delta$. The spectral weight of the three-magnon continuum at $q=\pi$ is $2.7\%$ compared to the single-magnon peak from SAC. These results are proved to be well matched with previous work \cite{PhysRevB.77.134437}. We have reason to believe that the elementary excitations of the Haldane phase are the bosonic magnons \cite{PhysRevB.78.094404}.

	The ground state of the topologically trivial large-$D$ phase includes the product state with $S^z=0$ at every site if the single-ion anisotropy $D$ is strong enough. Here, we choose an anisotropy $D/J=1.5$ for this phase. Predictably, the lowest-lying excitations can be viewed as single up or down spins that move in a background of ground state with $S^z=0$ \cite{PhysRevB.97.060403}. The quasiparticle excitations can be termed as excitons and antiexcitons, which reside in the $S^z=\pm1$ branch as shown in Fig. \ref{sqw}(c). Apparently, the large-$D$ phase also has an energy gap. A prominent peak can be observed near the wave vector $q=\pi$ and we consider it as single-exciton excitation. Besides, the extremely weak continua emerge at high frequencies similar to the Haldane phase (see Fig. \ref{spiw}). We suppose that
	they are multi-excitons and exciton-antiexciton bound states because of the interaction between opposite spins.

	Next, we focus on the quantum critical point that belongs to the TLL with a single-ion anisotropy $D/J\approx0.97$ in Fig. \ref{phase}. The accuracy of the critical point is high enough for the calculations of the dynamical spin excitations and therefore the single-ion anisotropy $D/J=0.97$ can be treated as the critical value of the quantum phase transition point. A symmetry-protected topological phase cannot be continuously changed into a trivial gapped phase without closing the energy gap, so we are easy to know the critical point is gapless. However, there is no well-defined quasiparticle excitations. Recently, the TLL phase of the $S=1/2$ chain has been realized experimentally, which verified that it possessed a broad continuum and its excitations are fractionalized spinons. From the spin excitation spectra shown in Fig. \ref{sqw}(b), gapless excitations appear at the wave vector $q=\pi$. Moreover, there is a broad continuum near $q=\pi$, which seems like the $S=1/2$ case. 
	
	Figure \ref{spiw} shows the transverse dynamic spin structure factor $S^{xx}(3\pi/4,\omega)$ and $S^{xx}(\pi,\omega)$ of the $S=1/2$ Heisenberg antiferromagnetic chain and the $S=1$ chain in the Haldane phase, the TLL critical state and the large-$D$ phase. We further find that the excitation spectra of the TLL state has a broader continuum than the Haldane phase and the large-D phase, meanwhile with the similar shape to the $S=1/2$ case, which has a high-frequency tail. To conclude, the quasiparticle excitations of the $S=1$ chain are spinon continuum excitations at the quantum critical point.

	\begin{figure}[t]
		\centering
		\includegraphics[width=8cm]{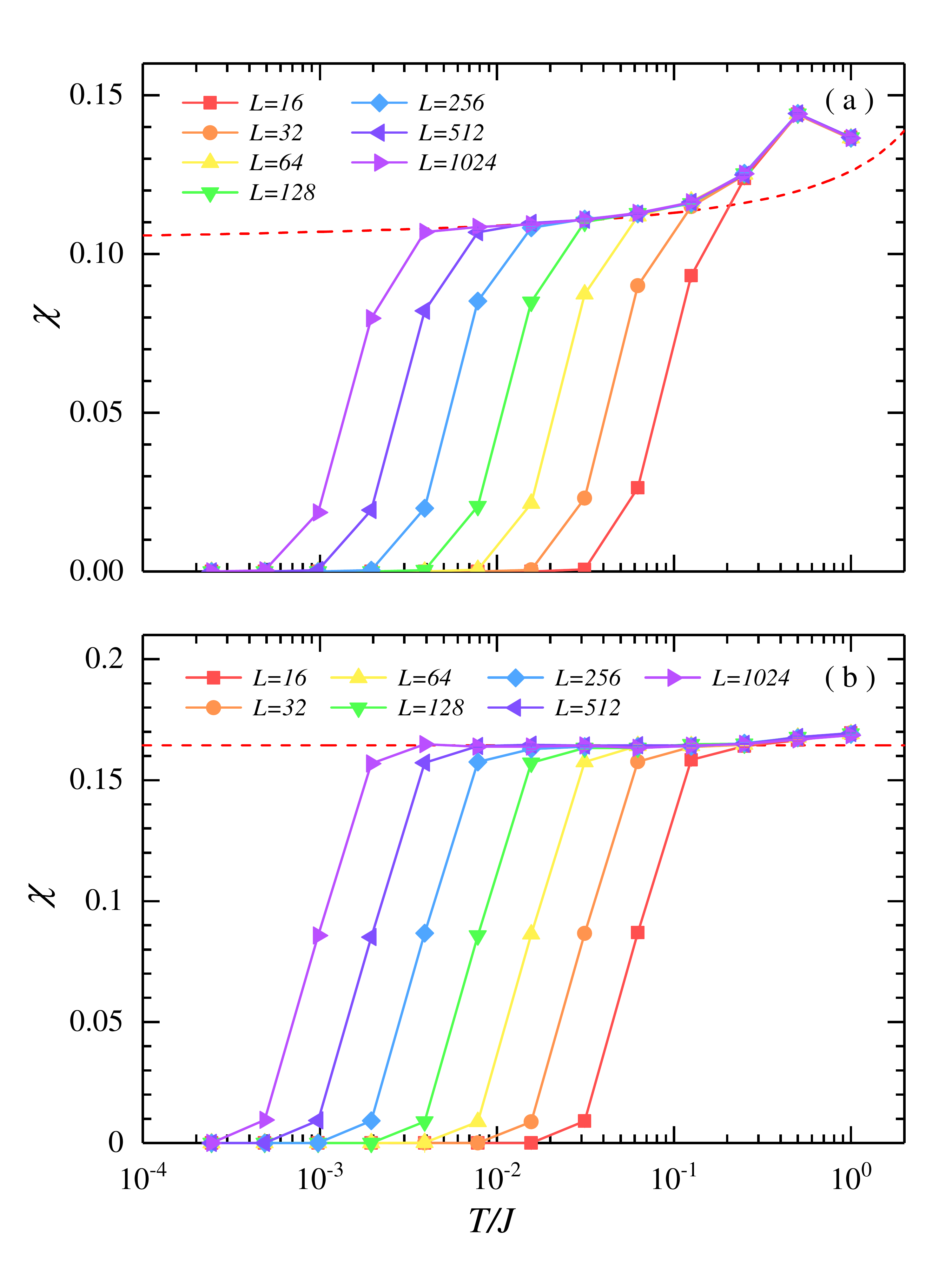}
		\caption{(Color online) (a) The uniform magnetic susceptibility of the $S=1/2$ Heisenberg chain. The red dashed curve is a fitting line based on Eq. (\ref{susceptibility}) with $v=\pi/2$ and $T_0=7.7$. (b) The uniform magnetic susceptibility of the $S=1$ Heisenberg chain in the TLL critical state. Both of them are obtained from the QMC calculations. The error bars are much smaller than the symbols.}
		\label{sus}
	\end{figure}
	
	\begin{figure}[t]
		\centering
		\includegraphics[width=8cm]{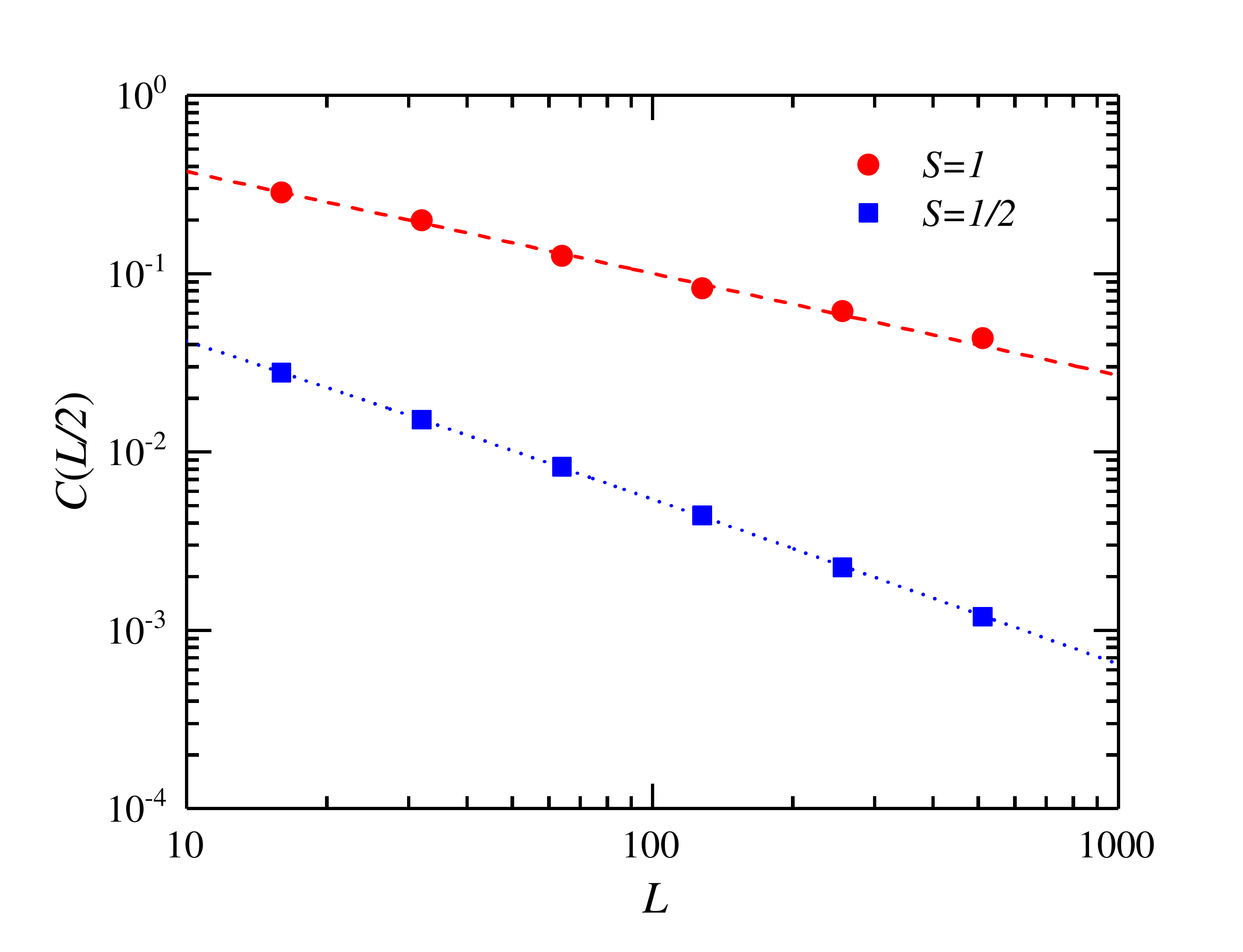}
		\caption{(Color online) The transverse spin-spin correlation function obtained from the QMC calculations for the $S=1$ Heisenberg chain in the TLL critical state and the $S=1/2$ Heisenberg antiferromagnetic chain with $L=16, 32, 64, 128, 256, 512$ and $\beta=1024$. The error bars are much smaller than the symbols. The red dashed line shows the form $\propto L^{-\alpha}$ with $\alpha=0.57(2)$ for $S=1$. We consider a logarithmic correction $ln^{1/2}(L/L_0)L^{-\alpha}$ for $S=1/2$ (blue dotted line) with $L_0=0.35(6)$ and $\alpha=1$.}
		\label{corr}
	\end{figure}
	
	\subsection{Uniform magnetic susceptibility}
	We further identify that the low-energy excitations are fractionalized spinons in the TLL critical state of the $S=1$ chain, which can be compared with a $S=1/2$ Heisenberg antiferromagnetic chain. From the low-energy field theory, the uniform magnetic susceptibility of the $S=1/2$ Heisenberg chain has the form \cite{PhysRevLett.73.332,PhysRevB.55.R3340}
	\begin{equation}
	\begin{aligned}
	\chi(T)=\frac{1}{2\pi v}+\frac{1}{4\pi v\ln(T_0/T)},
	\end{aligned}
	\label{susceptibility}
	\end{equation}
	where $v$ is the spinon velocity. As shown in Fig. \ref{sus}(a), the dashed curve is the low-$T$ form Eq. (\ref{susceptibility}) of the $S=1/2$ Heisenberg antiferromagnetic chain with $v=\pi/2$ and $T_0=7.7$ \cite{PhysRevLett.73.332}. The uniform magnetic susceptibility of the $S=1/2$ chain has been well studied before. We offer the details here in order to compare with the $S=1$ case. 
	
	In Fig. \ref{sus}(b), we presents the uniform magnetic susceptibility of the $S=1$ Heisenberg chain in the TLL critical state with the chain length $L=2^n$, where $n=4,5,...,10$. Because of the finite-size effects, the uniform magnetic susceptibility always decays to zero below a temperature $T~ (\sim1/L)$. For the $S=1$ chain, the log-linear scale makes the finite-size effects very clear and displays that the uniform magnetic susceptibility satisfies $\chi(T)\approx0.16439(9)$ in the low temperature as shown in Fig. \ref{sus}(b). This behavior is different from the $S=1/2$ chain. Thus, in the TLL state, the uniform magnetic susceptibility $\chi(T)$ of the $S=1$ chain will not alter with $T$ in the low temperature in the $L\to\infty$ limit, which is similar to the free fermion gas. The TLL critical state of the $S=1$ chain is paramagnetic and here we can regard the spinon excitations as bosons in this phase.
	
	\subsection{Transverse spin-spin correlation function}
	Additionally, we extract the transverse spin-spin correlation function $C(r)=\langle S^x_i\cdot S^x_{i+r}\rangle$ of the $S=1$ chain in the TLL critical state. We know the spin-spin correlation function $\langle S_i\cdot S_{i+r}\rangle$ of a $S=1/2$ Heisenberg antiferromagnetic chain has a power-law distribution as $(-1)^r/r$ while $C(r)$ of a isotropic $S=1$ chain decays exponentially with diatance $r$ as $(-1)^rr^{-1/2}e^{-r/\xi}$ \cite{PhysLettA93.464,PhysRevB.38.5188}. 
	
	In the TLL critical state, the transverse spin-spin correlation function $C(L/2)$ of the $S=1$ chain at the largest distance $r=L/2$ is shown versus the chain length $L$ and compared with a $S=1/2$ Heisenberg antiferromagnetic chain in Fig. \ref{corr}. For the $S=1$ chain, the transverse spin-spin correlation function has a power-law decay like the $S=1/2$ case, which is different from the Haldane phase. The decay of the $S=1$ chain is $\propto L^{-\alpha}$ with the exponent $\alpha=0.57(2)$ in the TLL critical state. For the $S=1/2$ chain, we consider a multiplicative logarithmic correction $ln^{1/2}(L/L_0)L^{-\alpha}$ with $\alpha=1$. The exponents of the $S=1$ and $S=1/2$ chains are different, but it is worth mentioning that they both have power-law correlations. Therefore, in the TLL critical state, the low-energy excitations of the $S=1$ Heisenberg chain are fractionalized spinons, just like a $S=1/2$ Heisenberg antiferromagnetic chain. Moreover, the $S=1/2$ chain has an emergent symmetry described by the SU(2) level-1 Wess-Zumino-Witten conformal field theory, which makes the spinon excitations available \cite{PhysRevB.93.115125,PhysRevLett.100.017203}. Nevertheless, at the topological quantum critical point of the $S=1$ Heisenberg chain, there is a U(1) spin rotational symmetry. They may belong to different universality classes and it is worthy of further research.

	\section{DISCUSSION AND CONCLUSION}		
	In this work, we have investigated the transverse dynamic spin structure factor $S^{xx}(q,\omega)$ of the $S=1$ Heisenberg antiferromagnetic chain versus the single-ion anisotropy $D$. We have uncovered the quantum phase diagram, comprising the Néel phase, the Haldane phase, the TLL critical state and the large-$D$ phase, especially their spin dynamics and elementary excitations. At the topological quantum critical point of topological phase transition between the Haldane phase and the large-$D$ phase, a broad continuum has been found near $q=\pi$ and precisely its dynamics are similar to a $S=1/2$ Heisenberg antiferromagnetic chain, which has been verified in various ways. So in the TLL state, the excitations of the $S=1$ Heisenberg antiferromagnetic chain also are fractionalized spinons.

	\begin{figure}[t]
		\centering
		\includegraphics[width=8cm]{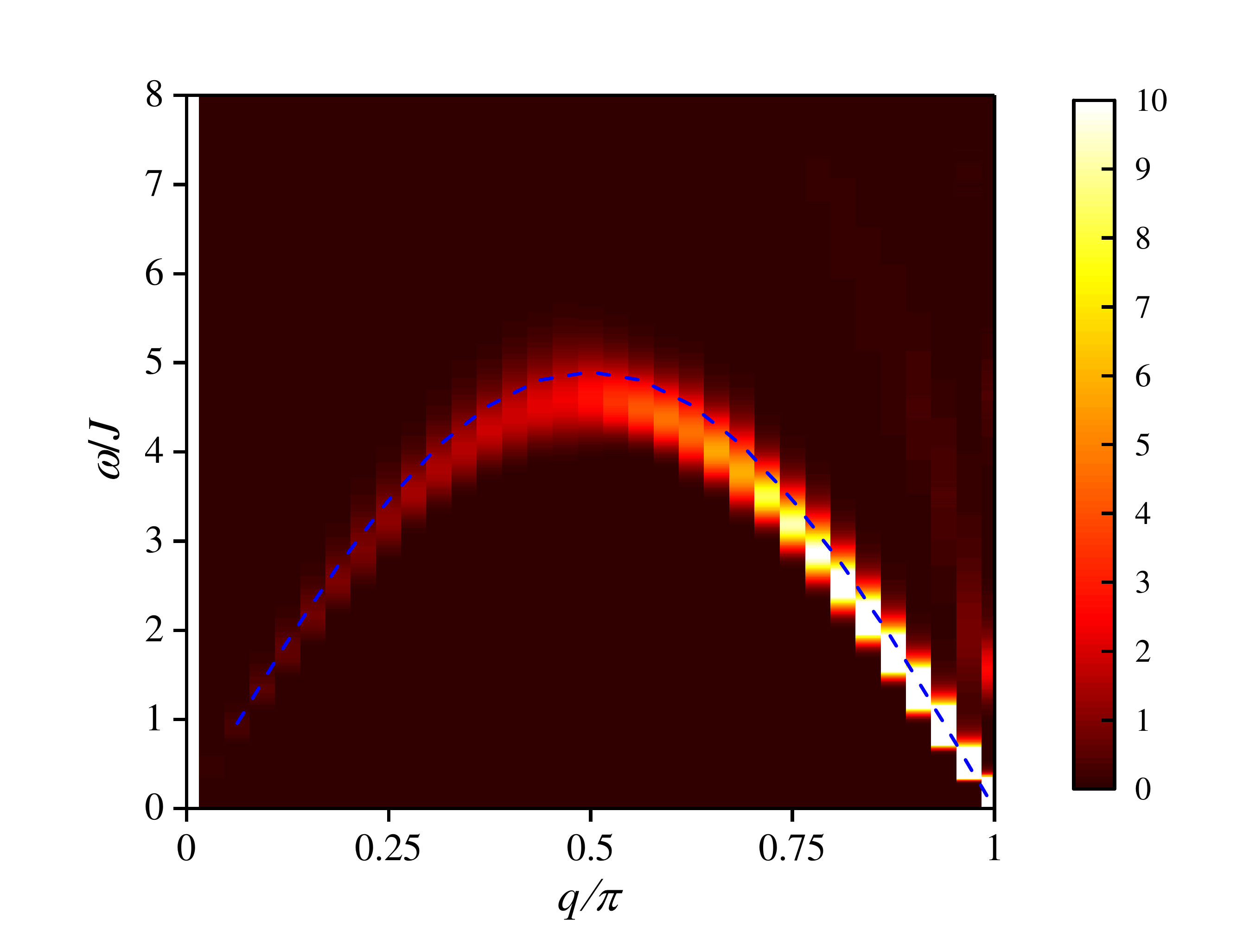}
		\caption{The dynamic spin structure factor $S(q,\omega)$ of the $S=2$ Heisenberg antiferromagnetic chain with $L=64$ and $\beta=16$. The blue dashed line is the magnon dispersion law with the form $\omega(q)=2J\sqrt{S(S+1)}sin(q)$ \cite{PhysRevB.48.6167}.}
		\label{spin2}
	\end{figure}

	Finally, we have extracted the dynamic spin structure factor of the $S=2$ Heisenberg antiferromagnetic chain \cite{PhysRevB.54.4038,EurophysLett30.493,PhysLettA213.102,PhysRevB.48.6167} as shown in Fig. \ref{spin2}. In contrast to the topological Haldane phase of the $S=1$ chain, the even-integral spin chain only has a topologically trivial gapped phase with a very small energy gap without a topological quantum phase transition \cite{PhysRevB.87.155114,PhysRevB.80.155131}. A prominent continuum of the $S=2$ chain can be observed and it seems to coincide with the previous results $\omega(q)=2J\sqrt{S(S+1)}sin(q)$ \cite{PhysRevB.48.6167}.

	\begin{acknowledgments}
		The authors would like to thank Anders W. Sandvik, Hui Shao, Nvsen Ma, Yu-Rong Shu and Yining Xu for helpful discussions. This work was supported by NKRDPC-2017YFA0206203, NKRDPC-2018YFA0306001, NSFC-11974432, NSFG-2019A1515011337, National Supercomputer Center in Guangzhou, Leading Talent Program of Guangdong Special Projects, and National Key Research and Development Program of MOST of China (2017YFA0302902).
		
	\end{acknowledgments}

	\bibliography{spin1}
	\bibliographystyle{apsrev4-1}
	
\end{document}